**Sub-THz and Hα activity during the preflare and main phases of a GOES class M2 event**

Pierre Kaufmann[1,2], Rogério Marcon[3,4], C. Guillermo Giménez de Castro[1], Stephen M. White[5], Jean-Pierre Raulin[1], Emilia Correia[1,6], Luis Olavo Fernandes[1], Rodney V. de Souza[1], Rodolfo Godoy[7], Adolfo Marun[7], Pablo Pereyra[7]

[1]CRAAM, Escola de Engenharia, Universidade Presbiteriana Mackenzie, São Paulo, Brazil

[2]CCS, Universidade Estadual de Campinas, Campinas, Brazil

[3]IFGW, Universidade Estadual de Campinas, Brazil

[4]Observatório Solar Bernard Lyot, Campinas, Brazil

[5]AFRL, Space Vehicles Directorate, Albuquerque, NM, USA

[6]Instituto Nacional de Pesquisas Espaciais, São José dos Campos, Brazil

[7]Complejo Astronômico El Leoncito, CONICET, San Juan, Argentina

**Abstract**

Radio and optical observations of the evolution of flare-associated phenomena have shown an initial and rapid burst at 0.4 THz only followed subsequently by a localized chromospheric heating producing an Hα brightening with later heating of the whole active region. A major instability occurred several minutes later producing one impulsive burst at microwaves only, associated with an M2.0 GOES X-ray flare that exhibited the main Hα brightening at the same site as the first flash. The possible association between long-enduring time profiles at soft X-rays, microwaves, Hα and sub-THz wavelengths is discussed. In the decay phase the Hα movie shows a disrupting magnetic arch structure ejecting dark, presumably chromospheric, material upwards. The time sequence of events suggests genuine interdependent and possibly non-thermal instabilities triggering phenomena, with concurrent active region plasma heating and material ejection.

## 1. Introduction

The time and space relationship between flares observed in Hα and associated bursts at X-ray and radio wavelengths can bring physical insights on the destabilization of an active region leading to the flaring processes. It is generally accepted that the soft X-ray bursts are due to heated plasma that is usually associated with flares identified as Hα brightening and radio bursts (Zirin 1988, and references therein). Evidence for such associations is dependent on the sensitivity of the measuring devices as well as on the time resolution of observations. On the other hand the association with sub-THz emissions is still poorly known for phenomena at low levels of activity.

The February 8, 2010, GOES M2.0 class solar burst (13:32-13:47-13:50 U.T.) provided a rare opportunity to compare an Hα flare image sequence obtained with moderate cadence (one frame every 10 seconds) to activity at sub-THz, microwaves and soft X-rays. We analyze flaring activity accompanying the burst in active region AR11045, located at 0 UT heliographic coordinates N23W01 (NOAA 2010), during the time period within about ± 15 minutes of the X-ray burst peak time (13:47 UT). We draw the attention to discrepancies between the AR and flare location reported by NOAA for this event, adopted here, and the reports from Solar Monitor (2010) and the SolarSoft list (2010).The latter two sources omitted the 13:32 U.T. event. The Solar Monitor reported an AR position for 8 February that corresponded to the location attained by 11045 on following day at 0 U.T (i.e. on 9 February). However all solar disk images, from all sources, confirm that the AR was near the central meridian at times close to 0 U.T. of 8 February.

Hα observations were obtained at "Bernard Lyot" Solar Observatory, Campinas, SP, Brazil (Marcon 2009). The 0.2 and 0.4 THz observations were obtained by the Solar Submillimeter Telescope (SST), San Juan, Argentina (Kaufmann *et al.,* 2010). Microwave observations were obtained at several frequencies by USAF RSTN monitoring network (Guidice, 1979) and by high sensitivity 7 GHz polarimeter observations at Itapetinga, Atibaia, Brazil (Kaufmann 1971; Correia *et al.* 1999).

## 2. The February 8, 2010 solar flare

We show in Figure 1 the AR1045 sunspot group in a SOHO MDI continuum solar image of the disk obtained at 12:10 UT (NASA 2010). The SST beams on the active region are shown with four 0.2 THz 4 arc minutes diameter beams (labelled 1-4) and two 0.4 THz 2 arc minutes diameter beams (labelled 5 and 6). Beams 2, 3, 4 and 5 are pointing at preselected coordinates of the active region, while beams 1 and 6 are 8 arc minutes off-source. The position of the sub-THz bursts relative to the SST antenna beams are inferred from the comparison of relative excess antenna temperatures measured at 0.2 THz by beams 2, 3 and 4 (Georges *et al.* 1989; Giménez de Castro *et al.* 1999). The positions are determined for every data point throughout the event duration. The position of the burst relative to the cluster of beams does not depend on the antenna pointing and tracking accuracy. This means that the antenna temperatures corrected for the source angular displacements with respect to the beams are independent of the accuracy of the pointing of the cluster of beams on the solar disk. The absolute uncertainty of the location of the cluster of the sub-THz beams on the solar disk is known to be better that 0.5 arc minutes (Kaufmann *et al.* 2010). This uncertainty is represented by the length of the cross arms in Figure 1, indicating a burst position at one time. The rectangular frame corresponds to the first Hα movie frame, obtained at 13:26:42 UT, shown in Figure 2. The angular scale of the field shown in Figure 2 is 126" x 174". Figure 3 shows time profiles at several wavelengths (X-ray, radio, Hα) together with a sequence of Hα images at significant flare phases (times as labeled). The top panel of Figure 3 shows the GOES 1-8 Å soft X-ray profile, exhibiting a small flux enhancement at about 13:33 UT followed by the main M2.0 burst starting at about 13:43 UT. The second panel from the top shows the high sensitivity 7 GHz total power flux on a log scale, starting with a small 2 SFU slow rise-and-fall event (1 SFU = $10^{-22}$ W/m$^2$ Hz) close in time with small enhancements at soft X-rays and Hα, followed by another rise-and-fall time structure (at about 13:42-13:14:00 UT) on which is superimposed an impulsive burst (at about 13:43:30-13:45:30 UT). The Hα time profile (i.e., the emission integrated over the entire field shown, in arbitrary units) is shown in the third panel from the top. The small short time scale fluctuations seen on the Hα profile are not real, and may be attributed to turbulent flow in the atmosphere during observations. The Hα seeing was poor for that day. In order to minimize the effects of seeing we have integrated the photon counts over a circular area containing about 15 pixels at the flaring site and subtracted it from the averaged data on an identical area near the flare.

The bottom two panels show the SST excess antenna temperatures at 0.2 and 0.4 THz corrected for atmosphere transmission and adjusted for the burst centroid position as deduced from the comparison of individual beams 2, 3 and 4 excess antenna temperatures. To minimize fluctuations due to signal propagation through the terrestrial atmosphere the 40 ms antenna temperature data at 0.2 THz from channels 2, 3, 4, were first smoothed over 50 points (or two seconds) and differenced with off-source channel 1 data, smoothed over 1000 points (40 seconds) to remove variability due to systematic effects such as atmospheric opacity and common receiver drifts. Similarly the 0.4 THz antenna temperature data from channels 5 was first smoothed over 50 points (or two seconds) and then differenced with off-source 40 ms channel 6 data, smoothed over 1000 points (40 seconds). The resulting solar antenna temperature contributions for channels 2, 3 and 4 are used to estimate the source position at 0.2 THz, shown as the cross in Figure 1. Once the source position is known the antenna temperature in each beam ca n be corrected for the offset of the source from the centroid of the beam. The resulting burst source antenna temperature time profile is plotted versus time in Figure 3 (bottom). The 0.4 THz data in channel 5 are corrected similarly, assuming that the burst positions are the same at both frequencies. Analysis of the burst centroid position versus time has shown that it remained stable at the location of the cross in Figure 1 (with respect to the orientation of the cluster of beams) to within less than ±30 arcseconds throughout the event duration.

One should keep in mind, however, that although the 0.2 and 0.4 THz time profiles shown in Figure 3 are corrected for long term atmosphere propagation effects, shorter time scales fluctuations remain more pronounced at 0.4 THz because of considerably poorer atmospheric transmission at the shorter wavelength. These fluctuations are not entirely cancelled by the differencing technique. Furthermore the differencing technique does not cancel eventual individual channel gain variations with time. Finally the resulting corrected antenna temperature has an uncertainty of order 25% due to the approximations made in the corrections of beam antenna temperature contributions (mainly due to atmospheric transmission, to uncertainty in the burst position with respect to the beam centroids, and to the determination of the baseline level needed to derive the excess temperatures).

The first sub-THz burst visible in the light curves for this event was detected at 0.4 THz only, appearing as a spike-like rapid burst, possibly complex, between 13:32:10 and 13:33:00 UT, at the beginning of the first slow rise in soft X-rays, 7 GHz and Hα, with excess antenna temperature of about 150 K, corresponding to about 93 SFU (with up to 25% uncertainty). The rapid complex burst is shown in expanded time scale in Figure 4, at 0.2 and 0.4 THz. The spike-like burst is easily distinguishable from all other 0.4 THz fluctuations throughout the time period studied here. The rapid burst signal time structures are at about a 6-10 sigma level.

Small and slow rise-and-fall time structures observed at 0.2 and 0.4 THz are also evident in Figure 3. One of them is seen between 13:43:30 and 13:50:10 UT, corresponding to excess corrected antenna temperature of about 60 K at both frequencies. The apparent good time correspondence of the sub-THz feature with the slow enhancements in GOES soft X-rays, 7 GHz and Hα suggests that this sub-THz rise is significant. The fluxes of the sub-THz small enhancements may be inferred if we assume sub-THz sources that have spatial sizes smaller than the SST beams (i.e., less than about 2 arc minutes). We obtain 37 and 21 SFU at 0.4 and 0.2 THz, respectively, which may correspond to an optically thick sub-THz thermal source. Another slower time structure at about 13:36-13:40 UT is observed at 0.4 THz, with an excess of about 50 K of corrected antenna temperature, but with less than about 20 K at 0.2 THz. However these slower time structures are too close to the detection limits to deserve further analysis.

White arrows in the selected Hα movie frames in Figure 3 indicate the principal features observed. The initial Hα localized flash sets up at about 13:32 UT. It was followed by a tenuous Hα E-W patch brightening with duration of order 10 minutes, as indicated in the frame at 13:41:42 UT. An important impulsive microwave burst occurred at about 13:43 – 13:46 UT, during the GOES soft X-ray rise time, with maximum emission at a frequency of about 7 GHz (based on USAF RSTN data from 1.4 to 15.4 GHz). However, no sub-THz counterpart of the microwave burst is seen by SST, to within an excess corrected antenna temperature limit of 5 and 20 K at 0.2 and 0.4 THz, respectively (i.e., about 2 and 13 SFU, respectively). The microwave burst coincided with the brightest phase of the Hα flare, which occurred at about the same location as the initial Hα flash. There are slow superimposed sub-THz fluctuations in the same time interval (13:43-13:53 UT), more pronounced at 0.4 THz, that might be real, although they are too close to the detection limits as discussed before. There were dynamic changes observed just after the brightest phase of Hα flare seen in the movie, followed by the sudden appearance of dark features over the southern part of the bright area, and in the region between there and the northern spot, as illustrated by white arrows on the movie frame at 13:52:42 UT, shown in Figure 3. These features suggest an early stage of the launch of dark material upwards: this material becomes more pronounced in later images (see frame at 14:00:42 UT in Figure 3).

### 3. Discussion

The microwave radio emission trends follow the general gradual-impulsive-gradual evolution that is often seen in flares (Kundu 1983). The principal event, reaching a maximum at about 13:44:30 UT, exhibited a rather typical flare relationship of the Hα emission to soft X-rays and microwaves as described qualitatively by Zirin (1988), accordingly to the relationship

Flare Importance = log (microwave flux) – 0.5

The Hα flare of importance 1 covered an area of about 300 millionths of the solar disc area, associated with an M2 soft X-ray burst and an impulsive microwave burst with peak flux of 100 SFU. This event was preceded by a small C-class GOES soft X-ray event correlated with a 2 SFU rise-and-fall microwave burst.

The observations shown here suggest that the whole energy conversion process started with a small Hα flash occupying < 100 millionths of the solar disc area, with undetectable microwave emission (if any) and rapid spikes seen only at 0.4 THz (at about 13:32 U.T). They might be associated with energetic phenomena and particle acceleration at the beginning of the whole process. There are several suggestions proposed to explain radio bursts that are only seen in the THz and sub-THz spectral domain (Sakai *et al.* 2006; Kaufmann & Raulin, 2006; Silva *et al.* 2007; Trottet *et al.* 2008; Fleishman & Kontar 2010). All of them require particles accelerated to very high energies, producing synchrotron spectra peaking somewhere in the THz range of frequencies. At the lower ≤ 0.2 THz frequencies the radio emission is reduced, either by the synchrotron emission mechanism (self absorption) and/or associated plasma effects (Razin suppression). Another alternative is that there might be a rapid heating process linked to the Hα fast flash, lower in the solar atmosphere, where the high electron density profile is such that it is optically thin at 0.4 THz and thick at 0.2 THz (as suggested for different scenarios, and slower processes, by Silva et al. 2005, and Trottet et al. 2011). In this particular case, the high energy process would be confined to a short interval of time, producing the sub-THz spike-like burst, probably occurring in a small region delimited by the first small Hα flash. They may have been responsible for the subsequent flare manifestations.

The subsequent long-enduring sub-THz fluctuations are too close to detection limits to be given much credence. However the antenna temperature enhancement occurred at about 13:43:30-13:50:10 UT coincides approximately in time with a slow-enduring component seen in the soft X-rays, the Hα time profiles and microwaves. The same excess temperature at 0.2 and 0.4 THz, of about 60 K, suggests emission from a dense optically thick thermal source. At the same time period the Hα movie suggests significant dynamic changes in the spatial structure. They might be related to processes that led to the dark material ejection observed immediately afterwards in Hα images.

The long enduring flare emissions at microwaves, soft X-ray and tentatively at sub-THz wavelengths might be compatible with the emission response of thermal plasma heated by the collision of high energy electrons with the dense active region ambient plasma (Lin & Hudson 1971). However, if we convert the SST antenna temperatures to fluxes – by assuming arbitrarily sources smaller than about 2 arcminutes - the SST fluxes become 37 and 21 SFU, at 0.4 and 0.2 THz respectively. These values are larger than the corresponding 7 GHz long-enduring component (~ 10 SFU). This result may suggest the presence of two thermal sources, one hot at microwaves and another cold at sub-THz, as suggested by Trottet *et al.* (2011), and (for quiescent active regions) by Silva *et al.* (2005). On the other hand, the impulsive microwave radio burst exhibiting peak spectral emission at about 7 GHz is typical for most of flares of this size and is probably caused by gyrosynchrotron emission of mildly relativistic electrons (Dulk 1985).

### 4. Concluding remarks

The sequence of events seen in this flare suggests that high energy particle acceleration can happen with just a small number of particles being involved, producing small fluxes in the sub-THz range and perhaps triggering other instabilities in the active region. These subsequent instabilities may accelerate particles at lower energies, producing emissions at microwave frequencies in conjunction with their Hα counterparts. The long-enduring burst at a later phase, observed at soft X-rays, Hα, microwaves and tentatively at sub-THz, and the Hα evidence for magnetic loop perturbation and disruption with upward ejection of dark material are possibly all part of the whole process.

This event demonstrates that solar activity diagnostics at sub-THz frequencies for low levels of activity are a useful tool to reveal the acceleration of high energy particles even when small in number, giving an insight into the high energy component that might also be present in flares of lesser Hα brightness importance.

**Acknowledgements.** The authors acknowledge the referee remarks that helped to improve considerably the paper presentation. This research was partially supported by Brazilian agencies FAPESP, CNPq and Mackepesquisa, Argentina agency CONICET and US agency AFOSR.

**Captions to the Figures**

Figure 1 - The AR11045 sunspot group in a SOHO MDI continuum solar image of the disk (NASA 2010) with the six SST antenna beams. The cross indicates the approximate position of the sub-THz bursts. The cross arm size of ±30 arcseconds corresponds to the SST pointing uncertainty The rectangle corresponds to the Hα movie frames (Figure 2).

Figure 2 - The Hα image obtained at 13:26:42 UT, showing the quiescent AR11045 before the flaring processes.

Figure 3 - The multiple-wavelength time profiles with selected Hα images. In the sequence of panels in the center from the top are plotted soft X-rays (GOES), showing the event beginning with a small enhancement followed by the M2 event; the 7 GHz time profile, showing a small precursor-like enhancement followed by a long enduring event on which an impulsive burst is superimposed; the Hα whole-frame light curve; the SST antenna temperatures at 0.2 and 0.4 THz corrected for atmospheric attenuation and for position with respect to the antenna beams, differenced with a beam pointing 8 arcminutes away from the active region. The spike-like burst at the beginning was observed only at 0.4 THz and is reproduced in greater detail in Figure 4. Other SST fluctuations are apparent ωιτη the principal one at time interval 13:44–13:53 UT (indicated by arrows in the bottom) as discussed in the text. Selected Hα images sampled at particular phases are shown at the corresponding times as labeled. From the left bottom frame in counterclockwise sense, with features indicated by white arrows: the quiescent AR; the single flash brightening; the E-W patch brightening; the brighter phase; the dark material showing up; and the arch evolving upwards in the solar atmosphere.

Figure 4 – Time-expanded time profiles of the corrected antenna temperatures at 0.2 and 0.4 THz, showing the spike-like initial burst only at 0.4 THz.

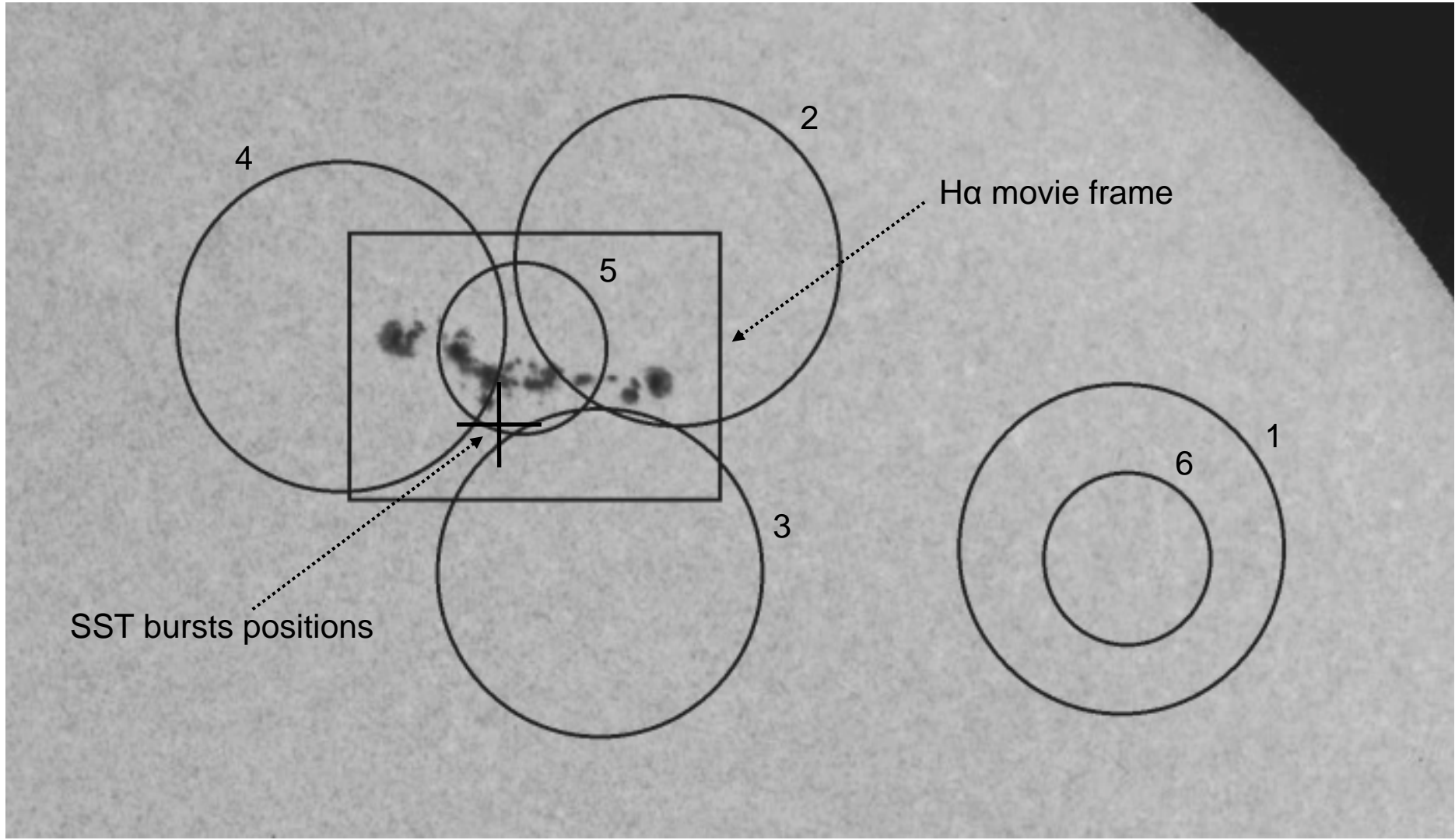

**Figure 1**

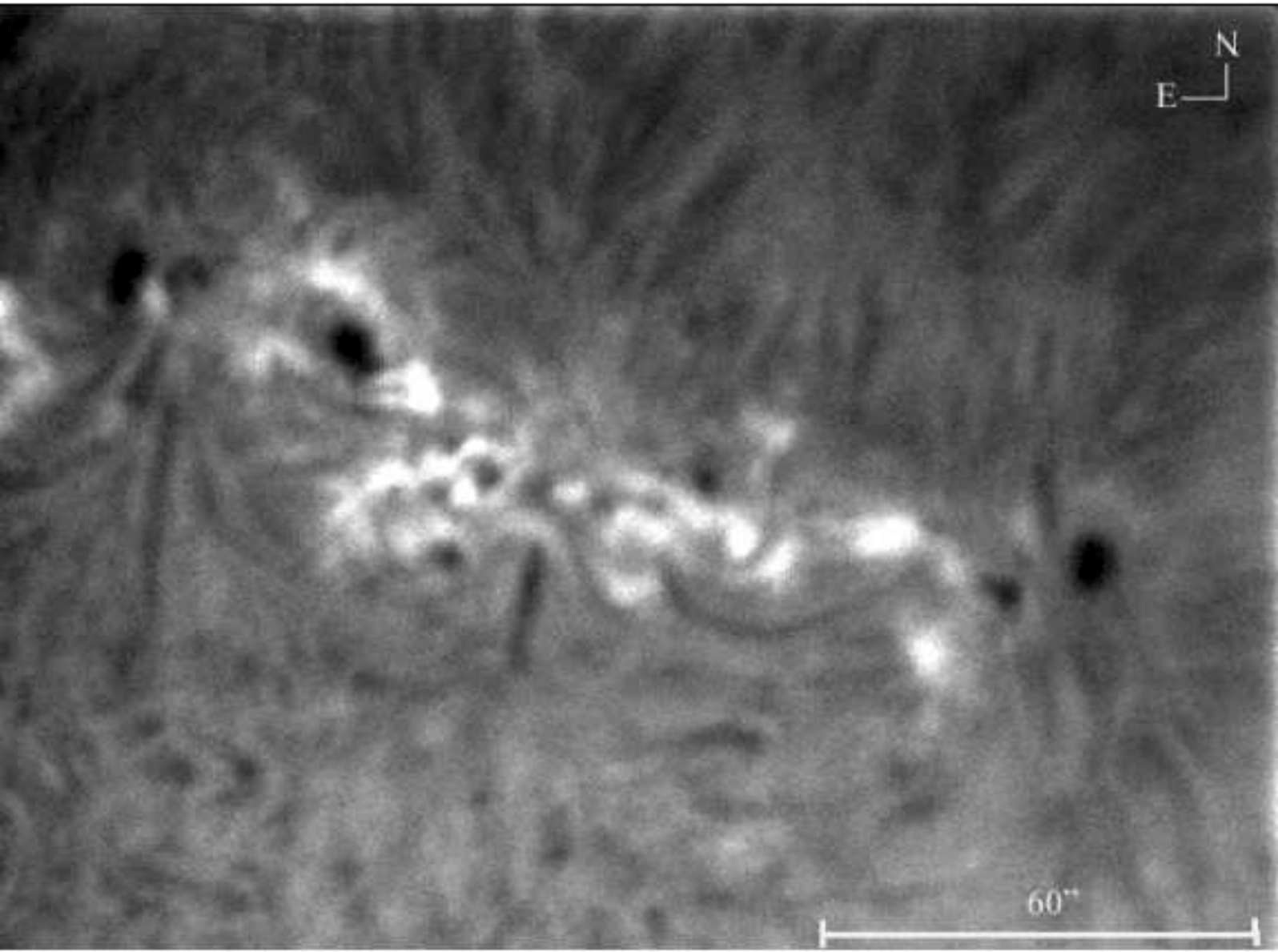

**Figure 2**

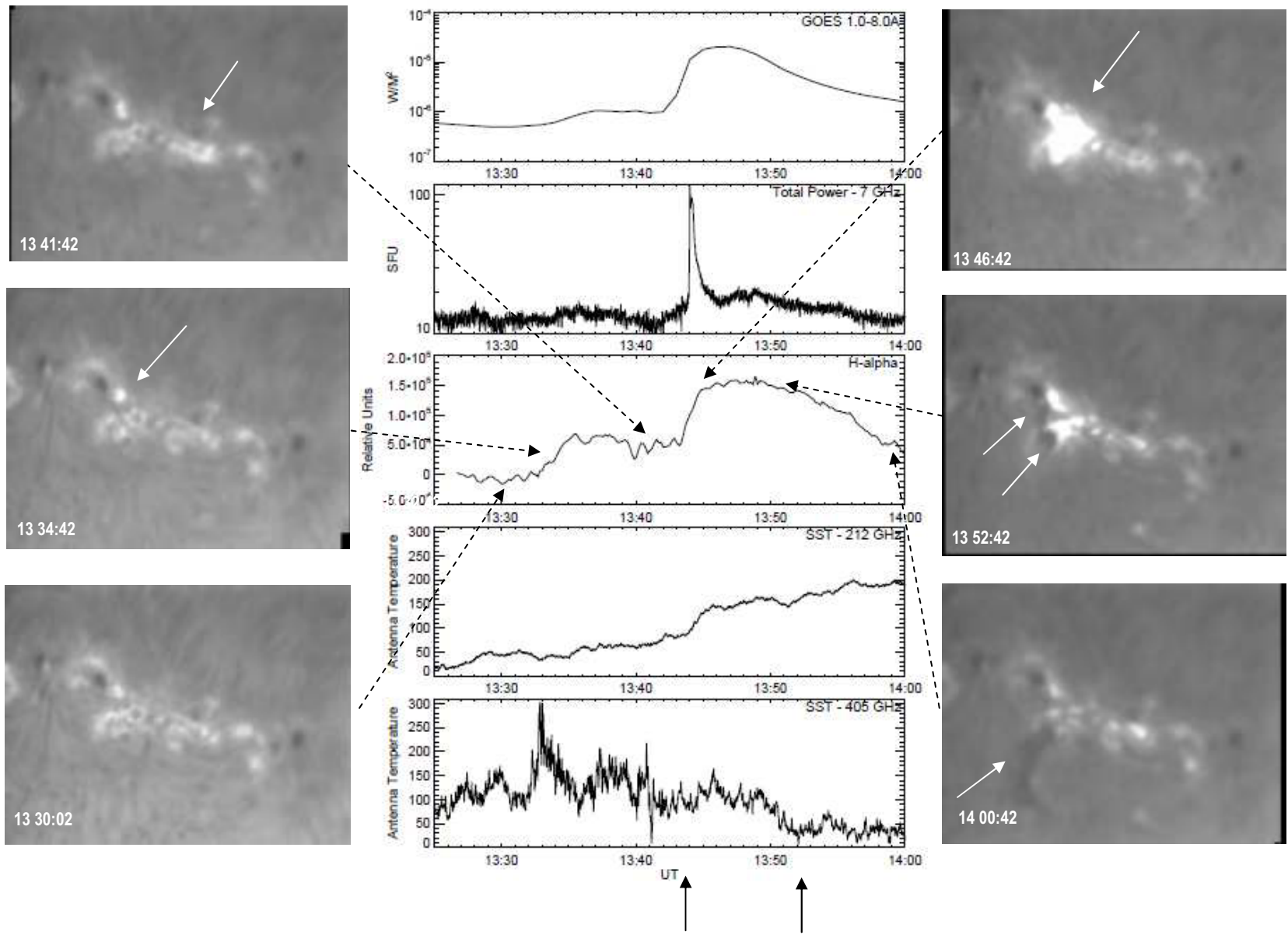

**Figure 3**

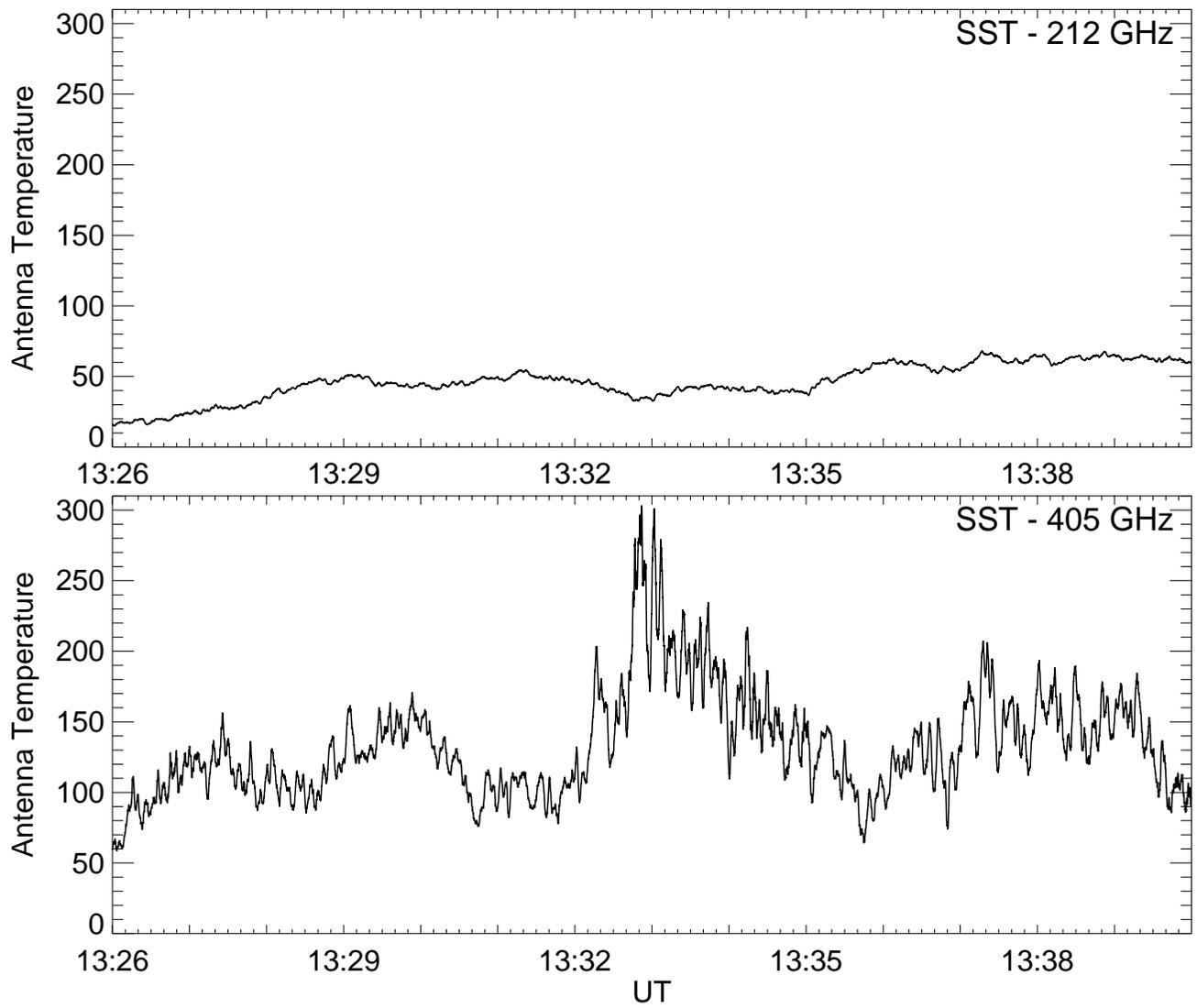

**Figure 4**